\documentclass[longauth]{aa}
\usepackage{txfonts}
\usepackage{graphicx}
\usepackage{longtable}
\usepackage{color}
\usepackage{multirow}
\usepackage{natbib}
\bibpunct{(}{)}{;}{a}{}{,}

\begin{document}

\title{Discovery of VHE $\gamma$--rays from the BL Lac object PKS\,0548$-$322}
\small {
\author{F.~Aharonian\inst{1,13}
\and A.G.~Akhperjanian \inst{2}
\and G.~Anton \inst{16}
\and U.~Barres de Almeida \inst{8} \thanks{supported by CAPES Foundation, Ministry of Education of Brazil}
\and A.R.~Bazer-Bachi \inst{3}
\and Y.~Becherini \inst{12}
\and B.~Behera \inst{14}
\and W.~Benbow \inst{1}
\and K.~Bernl\"ohr \inst{1,5}
\and A.~Bochow \inst{1}
% ...
}
}

\newpage

\institute {
Max-Planck-Institut f\"ur Kernphysik, P.O. Box 103980, D 69029
Heidelberg, Germany
\and
Yerevan Physics Institute, 2 Alikhanian Brothers St., 375036 Yerevan,
Armenia
\and
Centre d'Etude Spatiale des Rayonnements, CNRS/UPS, 9 av. du Colonel Roche, BP
4346, F-31029 Toulouse Cedex 4, France
\and
Universit\"at Hamburg, Institut f\"ur Experimentalphysik, Luruper Chaussee
149, D 22761 Hamburg, Germany
\and
Institut f\"ur Physik, Humboldt-Universit\"at zu Berlin, Newtonstr. 15,
D 12489 Berlin, Germany
\and
LUTH, Observatoire de Paris, CNRS, Universit\'e Paris Diderot, 5 Place Jules Janssen, 92190 Meudon, 
France
\and
IRFU/DSM/CEA, CE Saclay, F-91191
Gif-sur-Yvette, Cedex, France
\and
University of Durham, Department of Physics, South Road, Durham DH1 3LE,
U.K.
\and
Unit for Space Physics, North-West University, Potchefstroom 2520,
   South Africa
\and
Laboratoire Leprince-Ringuet, Ecole Polytechnique, CNRS/IN2P3,
F-91128 Palaiseau, France
\and 
Laboratoire d'Annecy-le-Vieux de Physique des Particules,
Universit\'{e} de Savoie, CNRS/IN2P3, F-74941 Annecy-le-Vieux,
France
\and
Astroparticule et Cosmologie (APC), CNRS, Universite Paris 7 Denis Diderot,
10, rue Alice Domon et Leonie Duquet, F-75205 Paris Cedex 13, France
\thanks{UMR 7164 (CNRS, Universit\'e Paris VII, CEA, Observatoire de Paris)}
\and
Dublin Institute for Advanced Studies, 5 Merrion Square, Dublin 2,
Ireland
\and
Landessternwarte, Universit\"at Heidelberg, K\"onigstuhl, D 69117 Heidelberg, Germany
\and
Laboratoire de Physique Th\'eorique et Astroparticules, 
Universit\'e Montpellier 2, CNRS/IN2P3, CC 70, Place Eug\`ene Bataillon, F-34095
Montpellier Cedex 5, France
\and
Universit\"at Erlangen-N\"urnberg, Physikalisches Institut, Erwin-Rommel-Str. 1,
D 91058 Erlangen, Germany
% ...
}




\date{Released 2009 Xxxxx XX}

\abstract
{}
{PKS\,0548$-$322 ($z=0.069$) is a ``high-frequency-peaked'' BL Lac object and a candidate very high energy (VHE, $\textrm{E}>100$\,GeV) $\gamma$-ray emitter, due to its high X-ray and radio flux. Observations at the VHE band provide insights into the origin of very energetic particles present in this source and the radiation processes at work.}
{We report observations made between October 2004 and January 2008 with the H.E.S.S.\ array, a four imaging atmospheric-Cherenkov telescopes. Contemporaneous UV and X-ray observations with the \emph{Swift} satellite in November 2006 are also reported.}
{PKS\,0548$-$322 is detected for the first time in the VHE band with H.E.S.S. We measure an excess of 216 $\gamma$-rays corresponding to a significance of 5.6 standard deviations. The photon spectrum of the source is described by a power-law, with a photon index of $\Gamma=2.86\pm0.34_{\rm{stat}}\pm0.10_{\rm{sys}}$. The integral flux above 200 GeV is $\sim 1.3\%$ of the flux of the Crab Nebula, and is consistent with being constant in time. Contemporaneous \emph{Swift}/XRT observations reveal an X-ray flux between 2 and 10\,keV of $F_{2-10\,\rm{keV}}=(2.3\pm0.2)\times10^{-11}\,{\rm erg}\,{\rm cm}^{-2}\,{\rm s}^{-1}$, an intermediate intensity state with respect to previous observations. The spectral energy distribution can be reproduced using a simple one-zone synchrotron self Compton model, with parameters similar those observed for other sources of this type. }
{}
\keywords{gamma rays: observations -- galaxies: active -- galaxies: BL Lacertae objects: individual: PKS\,0548$-$322}

\maketitle
%

\section{Introduction}
\label{intro}
BL Lacertae objects (BL Lacs) are an extreme class of Active Galactic Nuclei (AGN). The spectrum of these peculiar objects, which is  flat and associated with compact sources in the radio, extends up to the $\gamma$-ray band. None or weak emission lines are detected, and the radio and optical emission is highly polarized. BL Lacs are characterized by a rapid variability in all energy ranges, and display jets with apparent superluminal motions. The extreme properties of BL Lacs are explained by relativistic beaming, i.e. of a relativistic bulk motion of the emitting region towards the observer (see e.g. \citealt{Bland78, Urry95}).

The observed spectral energy distribution (SED) of BL Lacs usually shows (in a $\nu F_{\nu}$ representation) two spectral components. The first peak is located in the radio to X-ray range, whereas the second is at higher energies, sometimes in the VHE range. The SED is commonly explained by two different types of models. In leptonic models, the lower energy peak is produced by synchrotron emission of relativistic leptons in a jet that points towards the observer. The second peak originates in the ``inverse-Compton'' scattering of leptons off seed photons \citep{Ghi89}. Depending on the origin of the seed photons, the leptonic models are divided in two classes. In the ``synchrotron-self Compton'' (SSC) models, the seed photons come from the synchrotron photon field itself \citep{Marscher85}. In external Compton (EC) scenarios, the seed photons are provided by various sources, including the accretion disk and broad emission line regions \citep{Dermer93}. In hadronic models, the VHE emission is produced via the interactions of relativistic protons with matter \citep{Pohl00}, ambient photons \citep{Mann93}, magnetic fields \citep{Aha00,Muck01}. 
BL Lac objects are divided into classes defined by the energy of the synchrotron peak: ``low-energy-peaked'' BL Lacs (LBLs) have their peak in the IR/optical wavelength whereas ``high-energy-peaked'' BL Lacs (HBLs) peak in the UV/X-ray band \citep{Giom94,Pad95}. 

PKS\,0548$-$322 is a nearby ($z=0.069$, \citealt{Fosb76}) and bright BL Lac object hosted by a giant elliptical galaxy of absolute visual magnitude $M_V=-23.4$  \citep{Falo95,Wurt96} which is the dominant member of a rich cluster of galaxies. The synchrotron emission of this object peaks in the X-ray band and therefore it is classified as an HBL \citep{Pad95}. It is the third BL Lac detected by the HEAO X-ray satellite \citep{Mush78}. Since then, PKS\,0548$-$322 has been extensively studied by different X-ray experiments and satellites, showing a complex spectral behaviour. The X-ray spectrum deviates from a simple power-law \citep{Urry86}, and broken power-law models indicate a synchrotron peak energy in the 1-5\,keV range. \cite{Tueller08} reported the detection of PKS\,0548$-$322 in the hard X-ray band by \emph{Swift}/BAT. Absorption features reported  by \citet{Samb98}, were interpreted as the presence of circumnuclear ionized gas, but this has not been confirmed by other spectroscopic observations \citep{Blust04}. Thermal emission from the host galaxy on the kpc scale was detected by \emph{Chandra} \citep{Dona03}. More details about the X-ray history of this object can be found in \citet{Perri07}. The blazar was not detected in the MeV-GeV range by the EGRET detector \citep{Hart99}. PKS\,0548$-$322 is suggested as a candidate VHE emitter by \citet{Cost02}. Previous observations failed to detect $\gamma$-ray emission from this object. Atmospheric Cherenkov telescope (ACT) experiments set upper limits on the VHE $\gamma$-ray flux of this source (CANGAROO: $F_{E>1.5\,\rm{TeV}}<4.3\times10^{-12}\rm{cm}^{-2}\rm{s}^{-1}$, \citealt{Robe99}; Durham Mark VI telescope: $F_{E>300\,\rm{GeV}}<2.4\times10^{-11}\rm{cm}^{-2}\rm{s}^{-1}$, \citealt{Chad00}). The discovery of VHE $\gamma$-rays from PKS\,0548$-$322 with the H.E.S.S.\ Cherenkov telescopes is presented in this paper (see sections~\ref{hessobs} and~\ref{hessresults}). Contemporaneous observations were carried out in X-ray, UV, and optical (see section~\ref{swiftres}, ~\ref{uvotres} and ~\ref{atomres}) with the \emph{Swift} satellite. In section~\ref{sedres} a single zone SSC model is applied to the data and discuss observational prospects with \emph{Fermi}.


\section{H.E.S.S.\ Observations and Results}\label{hessobsresults}


\subsection{H.E.S.S.\ observations}\label{hessobs}
The H.E.S.S.\ array of atmospheric-Cherenkov telescopes \citep{Aha06a} observed PKS\,0548$-$322 between October 2004 and January 2008 with the full four-telescope array, for a total observation time of more than 63 hours.
After the application of quality-selection criteria and a dead-time correction, a total of 34.9 hours of good-quality data remain. The mean zenith angle of the observations is 10 degrees, which corresponds to a post-analysis energy threshold of $250\,\rm{GeV}$. The results hereafter are based on the H.E.S.S.\ 3D-model analysis (see \citep{Lem06}), where a 3D model is used to reconstruct the detected atmospheric shower induced by $\gamma$-rays. For each detected shower, the direction, energy and 3D-width are reconstructed.

$\gamma$-ray-like events are selected using cuts on image size, 3D-width, and telescope multiplicity. Only events that trigger at least three telescopes are kept in order to improve the $\gamma$-hadron separation. For the analysis, on-source data are taken from a circular region of radius $\theta$ around the source position, and the background is subtracted using the event background rate estimated with the ``ring background'' model \citep{Aha06a}, for which an annulus at the same distance to the center of the camera as the target position is used (excluding the region close to the source). The cuts applied for the analysis are summarised in Table~\ref{cuttable} and are taken from \citet{Tluc}.

\begin{table}[ht]\centering
\caption{\label{cuttable}Summary of applied cuts.}
\begin{tabular}{ll}\hline
 cut			&	value	\\\hline
 image centroid distance		&	2.5\,deg		\\
 image amplitude	&	$>$60\,photoelectrons	\\
 \# of telescopes	&	$\ge$\,3	\\	
 core impact position	&	$\le$\,500\,m	\\
 reduced 3D width $\omega$&	$<$0.002	\\
 $\theta^2$		&	$<$0.01\,deg$^2$\\\hline
\end{tabular}
\end{table}

\begin{figure}
\begin{center}
\includegraphics [width=0.48\textwidth]{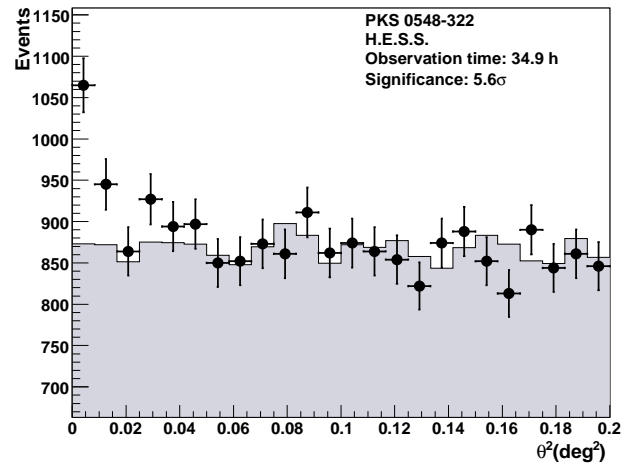}
\end{center}
\caption{Distribution of squared angular distance ($\theta ^2$) for counts from for on-source and normalized off-source events.}
\label{theta2_ring}
\end{figure}

\subsection{H.E.S.S.\ Results}\label{hessresults}
\begin{figure}
\begin{center}
\includegraphics [width=1.0\columnwidth] {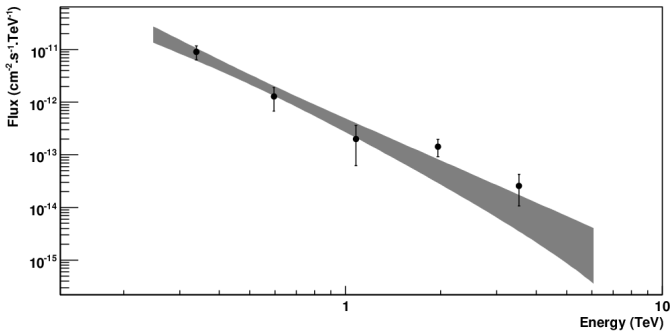}
\end{center}
\caption{VHE spectrum of PKS\,0548$-$322. The shaded region represents the 
1$\sigma$ confidence level of 
the fitted spectrum, using a power-law hypothesis. 
}\label{spectre}
\end{figure}

The distribution of squared angular distance ($\theta ^2$) for counts from the source is given in Figure \ref{theta2_ring}. A total of $N_{\rm{On}}$=1260 on-sources events and $N_{\rm{Off}}=3105$ off-source events are measured. The on-off normalisation factor is $\alpha=\Omega_{\rm{On}}/\Omega_{\rm{Off}}=0.336$, where $\Omega$ is the solid angle of the respective on- and off-source regions. The observed excess is $N_{\gamma}=N_{\rm{On}}-\alpha N_{\rm{Off}}=216$ $\gamma$-rays, corresponding to a significance of 5.6 standard deviations according to Equation~6 from \citet{Li83}. A two-dimensional Gaussian fit of the excess yields a position $\alpha_{\rm{J2000}}=5^{\rm{h}}50^{\rm{m}}38.4^{\rm{s}} \pm 3.2^{\rm s}, \delta_{\rm{J2000}}=-32^{\circ}16^{\prime}12.9^{\prime \prime}\pm 40.2^{\prime \prime}$. The measured position is compatible with the nominal position of PKS~0548$-$322 ($\alpha_{\rm{J2000}}=5^{\rm{h}}50^{\rm{m}}40.5^{\rm{s}}, \delta_{\rm{J2000}}=-32^{\circ} 16^{\prime}16^{\prime \prime}$) at the 1~$\sigma$ level. The differential photon spectrum shown in Figure \ref{spectre} is obtained using a forward-folding technique \citep{Piron}.  {Note that the likelihood maximisation provides the best set of parameters corresponding to a power law, with the corresponding error matrix. This is represented in Figure \ref{spectre} as a 68\% confidence-level band (grey zone). Points are derived from the residuals in different energy bins for illustration purposes only and plotted with 68\% confidence-level error bars in binned flux .}  
For a simple power-law hypothesis of the form $E^{-\Gamma}$, a likelihood maximization yields a spectral index of $2.86\pm0.34_{\rm{stat}}\pm0.10_{\rm{sys}}$ and an integral flux: I($>$ 250\,\rm{GeV} )= $({2.7\pm 0.6_{\rm{stat}}\pm 0.5_{\rm{sys}}}) \times 10^{-12}\,\rm{cm}^{-2}\,\rm{s}^{-1}$ ($\chi^2 /d.o.f.=24.2/14$). This corresponds to $\sim$1.3\% of the H.E.S.S.\ Crab Nebula flux (see \citealt{Aha06a}) above the same threshold, well below the upper limits reported by CANGAROO and the Durham Mark VI telescope (\citealt{Robe99,Chad00}). No evidence for flux variability is seen in the data:
{fitting a constant flux to the data, rebinned on a yearly timescale, gives a $\chi^{2}$ of 2.1/3 d.o.f. Furthermore, there was no 1 hour detection $>5\sigma$, which would be detected by H.E.S.S. if the source flux would have increased by a factor 4. Thus, a factor 4 of variability  of the source flux during more than 1 hour is ruled out, and a flux variability factor of $\mathbf{1+3/\sqrt{T}}$ is excluded during a time $T$ expressed in hours.}

\section{Multi-Wavelength Analysis and Results}

\subsection{Swift/XRT data reduction} \label{swiftres} 
Observations of PKS\,0548$-$322 were carried out by \emph{Swift} \citep{Gehrels04} contemporaneously with H.E.S.S.\ for an effective 4.3\,ks exposure on November 28, 2006 (obsId 0030836001). The XRT observation was performed in the Photon Count readout mode \citep{Hill04}. The XRT event files were calibrated and cleaned with standard filtering criteria with the \emph{xrtpipeline} task using the latest calibration files available in the \emph{Swift} CALDB distributed by HEASARC ({\it SWIFT-XRT-CALDB-11} released in May 2008). Events in the energy range 0.3--10\,keV with grades 0--12 was used in the analysis (see \citealt{Burr05} for a definition of XRT event grades). The source count rate was high enough to cause some photon pile-up in the inner 5 pixels ( $\sim 9"$ ) radius circle within the peak of the telescope Point Spread Function (PSF). Pile-up effect is avoided by selecting events within an annular region with an inner radius of 5 pixels and an outer radius of $\sim 20 $ pixels. The background is extracted from a nearby source-free circular region of 20 pixel radius. Ancillary response files for the spectral analysis are generated with the \emph{xrtmkarf} task, applying corrections for PSF losses ({\it psfflag=yes}). The PC mode grade 0-12 response matrix {\it swxpc0to12s0\_20010101v011.rmf} is used in the fits. The spectrum is binned to ensure a minimum of 50 counts per bin and the 0.3-10 keV recommended energy band is used. The flux is consistent with being constant during the observations. The spectral analysis is performed for the entire duration of the observation. A single power-law ($\Gamma=1.87\pm0.06$), with a Galactic absorption of $N_{\rm{H}}^{\rm{gal}}=2.52\times10^{20}\,{\rm cm}^{-2}$ consistent with the Galactic value \citep{Murphy96}, gives a reasonable fit ($\chi^2/{\rm d.o.f.}=40/25$). The integrated energy flux between 2 and 10\,keV is $F_{\rm{2-10\,keV}}=(2.3\pm0.2)\times10^{-11}\,{\rm erg}\,{\rm cm}^{-2}\,{\rm s}^{-1}$, an intermediate intensity state with respect to previous X-ray observations \citep{Blust04}. A broken power-law model with Galactic absorption yields a relatively smooth spectral break ($\Gamma_1=1.7\pm0.1$, $\Gamma_2=2.0\pm0.2$, $E_{\rm{break}}=1.7^{+1.0}_{-0.6}$\,keV). However, it is not a significantly better fit ($\chi^2/{\rm d.o.f.}=35/25$).
In both cases, i.e. with a power-law or a broken power-law, the possible presence of an absorption feature near 0.7 keV is observed. An F-test shows the addition of an absorption edged is significant at the 69\% and 92\% level for the power-law and broken-power-law cases, respectively. The presence of such an absorption feature is discussed in a recent paper by \cite{Blust04}. If real, it could be transient, as the analysis of XMM/RGS data by these authors reject the presence of such a feature in their spectra. In the present case, the significance is relatively weak.

{Although variability in optical, within a factor 2, has been reported for this source \citep{Perri07}, no clear indication of X-ray variability has been found in  \emph{Swift} observations, confirming Perri et al.'s findings. This would indicate that the source has been observed in a relatively steady state, although the existence of more active periods can not of course be excluded}

\subsection{Swift/UVOT data reduction}\label{uvotres}
The UltraViolet/Optical Telescope (UVOT) observations were made using 5 filter settings, during the same time slots as the XRT data.
Using the standard pipeline products (CALDB 20060917), the sequence of V, B, U, UVW1, UVM2 exposures have been checked for variability. No indication for variability is found and the individual frames are co-added. The reference stars are steady and UVOT photometry is within 0.03mag
consistent with the literature values \citep{poo08}. 
The chosen aperture is 3" radius for all filters to avoid contamination by foreground stars and decrease the background noise. An aperture correction,\footnote{http://heasarc.nasa.gov/docs/swift/analysis/threads/uvot\_thread\_ape rture.html} estimated to be of the order of $10\%$, has been added to the data. A similar sized region is used to estimate the background. 
The resulting count rate is corrected for Galactic absorption using a reddening of $\rm{E_{\rm{B-V}}^{\rm{gal}}}=0.035$ \citep{Schl98}, consistent \citep{Zombeck07} with the $N_{\rm{H}}^{\rm{gal}}$ from the power law fit  to the \emph{Swift}/XRT data. The color index is then converted into an extinction coefficient using the conversion factor provided by \cite{Giom06}. The extinctions are shown in Table~\ref{tab:extinction}.

\begin{table}[ht]\centering
\caption{Extinction coefficients for PKS\,0548--322 in the UVOT photometric bands.}
\label{tab:extinction}
\begin{tabular*}{0.8\linewidth}{@{\extracolsep{\fill}} lll}
\hline
Filter	&	$\lambda (\AA)$ & $A_{\lambda}$ (mag)	\\\hline
V	 			&	5460		&	0.112 	\\
B	   			& 	4350 	& 	0.147 	\\
U  			& 	3450 	&  	0.182 	\\	
UVW1		&	2600		&	0.2345 	\\
UVM2	&  2200	&  0.3395 	\\
UVW2		&	1930		&	0.2905	\\
\hline
\end{tabular*}
\end{table}
\subsection{ATOM data reduction}\label{atomres}
Optical observations were taken using the ATOM telescope at the H.E.S.S.\ site from November 2006 until January 2009 \citep{Haus04}. Absolute flux values have been calculated using differential photometry against two stars calibrated by \cite{Smith91} (B band) and \cite{Xie96} (V,R,I band). A 4'' radius aperture was used for all filter bands. 

A total of 312 measurements in 4 filter bands were taken in November 2006, May 2007, September 2007 to March 2008, and October 2008 to January 2009. Within errors, the photometry is compatible with a constant flux value of $m_B=17.0$, $m_V=16.2$, $m_R=15.6$, and $m_I=14.9$ ($4.4 / 6.7 / 8.1/ 10.4\,\times10^{-12}\,\textrm{erg}\,\textrm{cm}^{-2}\,\textrm{s}^{-1}$) for the B, V, R and I band. It is compatible with the assumption that the optical emission comes dominantly from the host galaxy. B and V band are systematically 0.2 mag brighter with respect to UVOT values, but we consider it consistent within the instruments systematic errors \citep{poo08}. 

\section{Spectral Energy Distribution and discussion}\label{sedres}
\begin{table*}[ht]\centering
\caption{Parameters used to fit the data (see text for further details).}
\label{tab:param}
\begin{center}
\begin{tabular*}{0.8\textwidth}{@{\extracolsep{\fill}} lcccccccccc}
\hline
	& $\delta$ & $R$ & $B$ & $K$ & $u_{e}/u_{B}$ & $\gamma_{\rm{min}}$ & $\gamma_{\rm{break}}$ & $\gamma_{\rm{max}}$ & $n_{1}$ & $n_{2}$ \\
	&  &  $[10^{16}\,\rm{cm}]$ & [G] & $[\rm{cm}^3]$ & & & &  & &\\
	\hline
 & $10$ & $1.9$ & $0.1$ & $2.6\times10^3$ & $30$ & $1$ & $2.7\times10^5$ & $5\times10^6$ & $2.2$ & $3.0$  \\ 
\hline
\end{tabular*}
\end{center}
\end{table*}

\begin{figure*}[htb]
\includegraphics[width=0.9\textwidth]{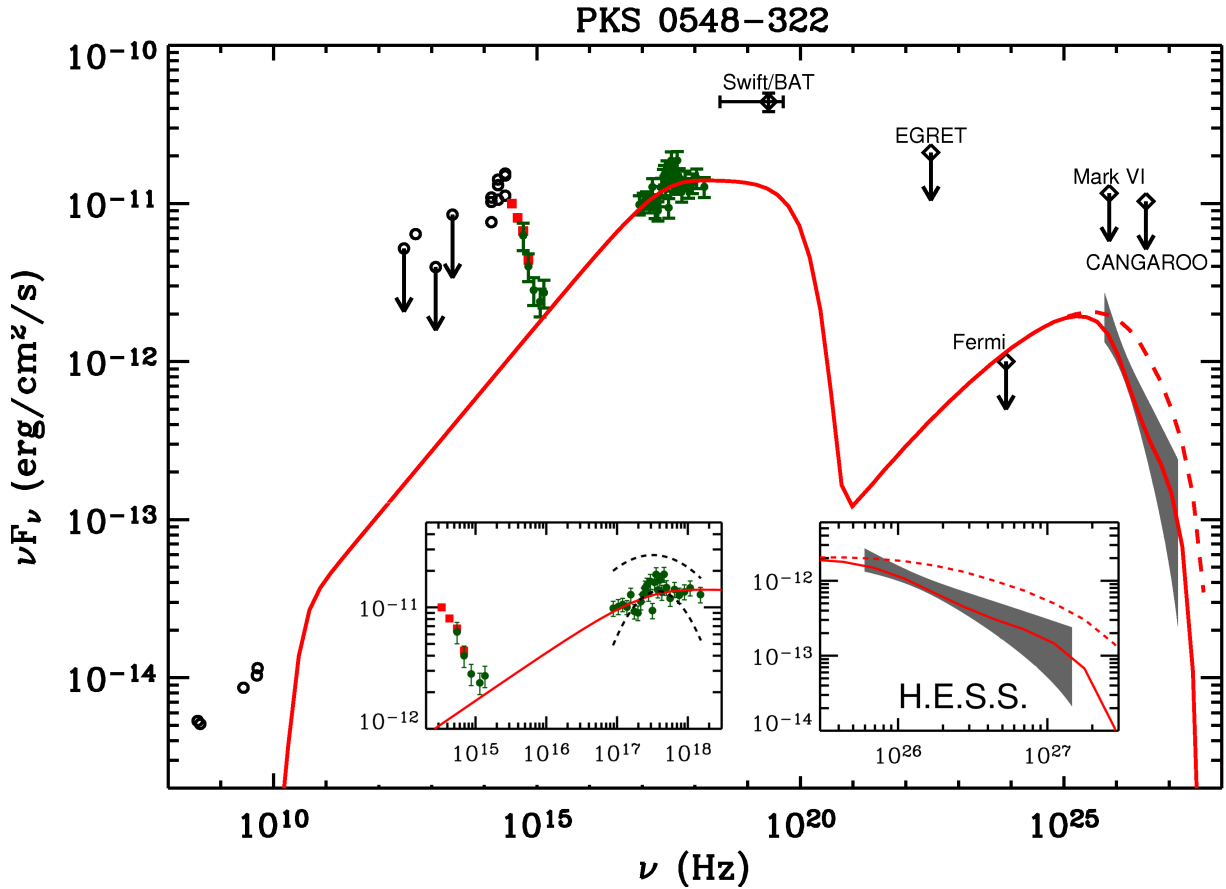}
\caption{Spectral energy distribution of PKS\,0548$-$322. Open circles: Radio, IR, and optical archival data are from the NED database. Fill red squares: ATOM data. Fill green circles: $\emph{Swift}$ observations contemporaneous with H.E.S.S.\ observations (corrected for Galactic absorption). Open diamonds: upper limits from \emph{Swift}/BAT, EGRET, Mark VI, and CANGAROO. The shaded region represents the $1\sigma$ confidence bounds of the spectrum observed by H.E.S.S.\ above 250\,GeV.  The dotted lines at VHE show the spectrum of the source de-absorbed of the EBL attenuation. The left-hand side inlay details a portion of the observed UV to X-ray spectrum. The dotted lines are the low level and high level flux observed with \emph{Swift} and reported in \citealt{Perri07}. The right-hand side inlay details the H.E.S.S.\ spectrum. The upper limits deduced from first year \emph{Fermi} observations \citep{Abd09} are also reported.}
\label{figure_SED_0548}
\end{figure*}

Figure~\ref{figure_SED_0548} shows the SED of PKS\,0548$-$322. Radio, IR, and optical archival data are taken from the NED\footnote{The NASA/IPAC Extragalactic Database (NED) is operated by the Jet Propulsion Laboratory, California Institute of Technology, under contract with the National Aeronautics and Space Administration.} database. X-ray data from \emph{Swift}/XRT are shown corrected for Galactic absorption, using the energy dependence for the photoelectric cross-section provided by XSPEC. The \emph{Swift}/UVOT data are corrected for Galactic absorption using the extinction coefficients in Table~\ref{tab:extinction}. The \emph{Swift}/BAT flux level in the 14--$195\,\rm{keV}$ range reported by \cite{Tueller08} is interpreted as an upper limit in the SED, and this for several reasons. First, this flux is higher by a factor of 2 compared to the 2--$10\,\rm{keV}$ XRT flux, and is not consistent with the spectral index after the energy break found in the \emph{Swift}/XRT analysis. Furthermore, these hard X-ray observations were integrated over 9 consecutive months that are not contemporaneous with the H.E.S.S.\ observations, and variability cannot be ruled out. Secondly, \cite{Tueller08} use a spectral index of 2.15 for all their sources. Using the BeppoSAX fit \citep{Perri07}, the index would be 2.3-2.8 in the energy range 14--$195\,\rm{keV}$. Hence, the harder assumption made by \cite{Tueller08} probably overestimates the actual flux in that band. The upper limits from \emph{EGRET} \citep{Hart99}, CANGAROO \citep{Robe99} and Mark VI Durham telescope \citep{Chad00} are also displayed. The $1\sigma$ confidence level of the VHE spectra measured by H.E.S.S.\ is shown, as well as the one year \emph{Fermi} upper limits \citep{Abd09}, estimated at the decorrelation energy (where the upper limit doesn't depend on the assumed spectral index). 

A simple one-zone, homogeneous, time independent, synchrotron self-Compton (SSC) leptonic model is applied to interpret the contemporaneous observations by H.E.S.S., \emph{Swift}, and ATOM. The radio data is assumed to originate from an extended region farther along the jet and the bump in IR-optical is attributed to emission from the host galaxy. Hence, these data are not fitted by the model. 

The SSC model describes the system as a spherical emitting region of radius $R$, filled with a tangled magnetic field $B$, and propagating with a bulk Doppler factor $\delta$ (e.g. \citealt{Band85, Katar01, Gie07}). The energy distribution function (EDF) of the radiating leptons is described by a broken power-law, with indices $n_1$ and $n_2$, between Lorentz factors $\gamma_{\rm{min}}$ and  $\gamma_{\rm{max}}$, with a break energy $\gamma_{\rm{break}}$ and a normalisation factor $K$. The absorption by the extra-galactic background light (EBL) is taken into account, adopting the model of \cite{Franceschini08}. The luminosity distance is $d_L=9.6\times10^{26}\,{\rm cm}$, assuming the following cosmological parameters: $H_{0}=70\,\rm{km}\,\rm{s}^{-1}\,\rm{Mpc}^{-1}, \Lambda=0.7, \Omega=0.3, q_{0}=-0.5$.

The model parameters are underconstrained, and many different  solutions provide a satisfactory fit of both X-ray and VHE spectrum. We fit the data assuming that the flux did not exceed the (non-contemporaneous) \emph{Fermi}upper limit at the time the H.E.S.S. observations occured. The parameters used in the model are summarised in Table~\ref{tab:param}. Data can be fitted with a reasonable  bulk Doppler factor $\delta=10$. A lower value would result in a significant internal $\gamma-\gamma$ absorption. The EDF is a broken power-law with indices $n_{1}=2.2$ and $n_{2}=3$, constrained by the shape of the synchrotron spectrum. These values are compatible with the index of the broken power-law that fits the X-ray data from \emph{Swift}/XRT. We emphasize that the value of the break $\Gamma_2-\Gamma_1$ observed by \emph{Swift} is compatible (within errors) with 0.5, and the dominance of the synchrotron component compared to the Compton one is consistent with a relativistic plasma losing its energy primarily via synchrotron cooling. The EDF is defined between $\gamma_{\rm{min}}=1$ and $\gamma_{\rm{max}}=5\times10^6$. The value of the maximal Lorentz factor $\gamma_{\rm{max}}$ is not well constrained due to the lack of contemporaneous observation in the hard X-ray range. The break energy in the EDF is associated with the synchrotron break that is detected around $4\times10^{17}\,\rm{Hz}$. In the model, the energy break depends on the magnetic field of the emitting region with the following relation: $\gamma_{\rm{break}}\propto (B\delta)^{-1/2}$, and is thus well constrained for a given Doppler factor. By choosing a value for the magnetic field $B$, the two last parameters of the model ($K$ and $R$) are constrained by the flux in X-ray and at VHE. 

With the adopted parameters, the emitting region is out of equipartition with a ratio of the particle energy density to the magnetic energy density  $u_{e}/u_{B}=30$. However this high value is in agreement with that usually derived in VHE blazars using one-zone models. 
One can also compute from these parameters a synchrotron cooling timescale $t_{\rm{cool}}\sim280\,\rm{ks}$ at the energy break, smaller than the source crossing time $t_{\rm{cross}}=R/c =640\,\rm{ks}$. Both values must be divided by $\delta =10$ to give an apparent dynamical timescale $t_{\rm{var}}=R/\delta c \sim 64\,\rm{ks}$, and an apparent cooling timescale $\sim28\,\rm{ks}$ , both being less than one day. The absence of a clearly detected short term variability does not allow to put a strong constraint on these parameters, although historical observations do exhibit some variability at year scale  \citep{Blust04}, which is clearly compatible with these values. The radius of the emitting source derived from the model is about a few hundreds times the gravitational radius of the blazar's central black hole, assuming a mass of $M_{\rm{bh}}\sim10^{8.5}M_{\odot}$ \citep{Barth03}. A larger Doppler factor could give an equally good fit with a smaller radius, a smaller break Lorentz factor, a larger B field, but a higher ratio of particle to magnetic energy density, so more out of equipartition. However these solutions would be favored if shorter variability timescales were detected. In any case, the low Compton to synchrotron flux ratio indicates that the source is (on average ) far from the Compton catastrophe, although more intense, short-lived flares can not be excluded.


\section{Conclusion}
Observations performed by H.E.S.S.\ from 2004 up to 2008 have established PKS\,0548$-$322 as a VHE $\gamma$-ray source amongst the closest blazars. The VHE flux is consistent with being constant within the H.E.S.S.\ observation period. The contemporaneous X-ray data can be fit by a broken power-law model with Galactic absorption.\\ 
For the first time for this source, a SED comprising contemporaneous optical, UV, X-ray and VHE measurements is made. A one-zone SSC model, taking into account absorption by the EBL, provides a satisfactory description of these data. 
Observations of this object with \emph{Fermi} should help to place stronger constraints on the model parameters by building a more complete SED of the object.
It is a further confirmation that the proportion of TeV sources among close blazars is fairly high. The Doppler factor used in the modeling by a simple one-blob model is compatible with the value commonly adopted for relativistic jets, although, as is often the case for TeV sources, no superluminal motion has been reported for this object. The addition of new objects to the set of ``high-frequency-peaked'' BL Lacs observed at VHE, combined with Fermi data, should soon enable population studies and insights into the underlying physical processes.

\begin{acknowledgements}
The support of the Namibian authorities and of the University of in facilitating the construction and operation of H.E.S.S.\ is gratefully acknowledged, as is the support by the German Ministry for Education and Research (BMBF), the Max Planck Society, the French Ministry for Research, the CNRS-IN2P3 and the Astroparticle Interdisciplinary Programme of the CNRS, the U.K. Science and Technology Facilities Council (STFC), the IPNP of the Charles University, the Polish Ministry of Science and Higher Education, the South African Department of Science and Technology and National Research Foundation, and by the University of Namibia. We appreciate the excellent work of the technical support staff in Berlin, Durham, Hamburg, Heidelberg, Palaiseau, Paris, Saclay, and in Namibia in the construction and operation of the equipment.

This research has made use of the NASA/IPAC Extragalactic Database (NED) which is operated by the Jet Propulsion Laboratory, California Institute of Technology, under contract with the National Aeronautics and Space Administration.

The authors acknowledge the use of the publicly available \emph{Swift} data, as well as the public HEASARC software packages.

\end{acknowledgements}

\newcommand{\thc}{(\hess\ Collaboration) }

\end{document}